\documentstyle[prl,twocolumn,aps]{revtex}

\begin{document}
\draft

\preprint{Submitted to PRL March 9, 1994}
\title{
EFFECTIVE MASSES OF IONS IN SUPERFLUID $^3$He-B}
\author{N. V. Prokof'ev}
\address{
Physics Department, University of British Columbia, 6224 Agricultural
Rd., \\
 Vancouver B.C., Canada V6T 1Z1 \\
and \\
 Kurchatov Institute, 123182 Moscow, Russia}
\maketitle

\begin{abstract}
We show that ion masses in  superfluid $^3$He ought to be enormously enhanced
(by a factor of $10^2$) as compared with the same  ion masses in
$^4$He measured
at low temperature. We calculate precisely the dependence of the effective
mass on pressure in  $^3$He-B, and show that the coherent (ballistic) motion of
ions in $^3$He-B can be
studied experimentally at $T<(0.3-0.2)\:T_c$.
\end{abstract}


The problem of ion motion in normal $^3$He-liquid has been of long-standing
interest, partly because of its connection with the
"orthogonality catastrophe", but mostly because theorists have had a
hard time explaining it. The ion motion is greatly overdamped at low
temperature, by multiple scattering of $^3$He quasiparticles, so theorists
have concentrated on calculating the experimentally measurable ion mobility.
Early perturbative calculations \cite{Abe} predicted a mobility $\mu (T)$
diverging as $1/T^2$ below a temperature $T_0=p_F^2/M$, where $M$ is the
bare ion effective mass ( $M \sim 100 - 260 m_3$, depending on pressure,
for the negative ion; here $m_3$ is the $^3$He atomic mass). Experiments on
both positive \cite{And68,Roach,Alex} and negative
\cite{And68,Kuchnir,McClintock,Ahonen,Nummila} ions flatly contradicted
this prediction; $\mu_- $ is roughly constant through and below $T_0$,
all the way down to the superfluid transition $T_c$.

However this problem is a strong-coupling one. The dimensionless ion-$^3$He
coupling is $g= p_F^2\sigma_{tr} / 3\pi^2$, with $ \sigma_{tr}$
 the transport cross section, and $g \gg 1$. The high-T scattering rate
equals $\Gamma =T_0 \:g\gg T_0$, which is why the ion motion is overdamped
already for $T\gg T_0$. Moreover, it was realised by Josephson and Lekner
\cite{Joseph} that for $T<\Gamma $ the ion recoil is not free, but Brownian
diffusive, down to the unobservably low temperature $T_{coh} =T_0 \:ge^{-g}$.
This diffusive motion means that it is meaningless to define an effective mass
for the ion above $T_{coh}$. The theory of ion mobility in normal $^3$He has
nevertheless been considerably refined since then \cite{Bow,Chen,Ya}.

One obvious way for experimentalists to see coherent motion of an ion in
$^3$He is to go to the superfluid phases, where the gap cuts off the
"orthogonality catastrophe". Remarkably, this possibility has not been
explored, either in theory or experiment (although some mobility
experiments have been done \cite{Nummila}- we return to these below). In this
paper I give a detailed theory of ion dynamics, which is exact in the
large-$g$ limit. A very striking prediction emerges from this analysis -
that the effective mass of ions in the superfluid phases will be very large
(up to $2\cdot  10^4 \:m_3$, or some 100 times the bare ion mass). I
calculate  the effective mass $M^{eff} (P)$ as a function of pressure in
the low-T limit in  $^3$He-B, and suggest how this prediction might be
verified experimentally. This prediction (which is  clearly
out of the framework of the standard models \cite{Atk}) should constitute a
very stringent test of our ideas of particle dynamics in a Fermi liquid.

The Hamiltonian  is that of a spherical object in a Fermi
liquid environment:
\begin{eqnarray}
H&=&{1\over 2}M\dot{\bf R}^2 + H_F + V\;,
\nonumber \\
V&=&\int d{\bf r}\;V({\bf r}-{\bf R}){\hat \rho }({\bf r}) \;,
\label{4}
\end{eqnarray}
where ${\bf R}$ is the ion coordinate, $H_F$ the  Hamiltonian of $^3$He,
and ${\hat \rho }({ \bf r})$  the $^3$He density operator.
We make  use of the path integral
technique and  integrate out the Fermion degrees of freedom
\cite{Chen,Ya}, and start by  considering the case of normal
 $^3$He.
Using Feynman's path integral over ${\bf R}$ in imaginary time \cite{Feynman}
the effective action in the partition function can be written
\begin{equation}
S=S_0-\!
\int\!\!\!\!\int_{0}^{\beta}\!d\tau d\tau ^\prime\:
{\pi\over \beta}
\sum_{n\ne 0}{\cal  F}_n({\bf R}_{\tau}\!-\!{\bf R}_{{\tau}^\prime})
e^{i\omega_n(\tau -\tau^\prime)},
\label{5}
\end{equation}
where $S_0=\int d\tau\:M\dot{\bf R}^2/ 2 $, $\omega_n=2\pi n/\beta $ are
Matsubara frequencies, and the influence
functional, ${\cal  F}_n({\bf R})$, is related to the overlap
integral between the initial and final Fermi liquid states with different
local potentials \cite{Ya}; one has
\begin{equation}
{\cal  F}_n({\bf R})={\mid \omega_n\mid \over 16\pi^2}\:Tr\{
\ln^2(S_fS^{-1}_i)\} \;.
\label{6}
\end{equation}
 Here $S_f=S({\bf R})$ and $S_i=S(0)$ are the scattering
$S$-matrices  at the Fermi energy in the final (the particle at the point
 ${\bf R}$) and initial ( ${\bf R}=0$) states. The connection with the
overlap integral $ \langle f\: \mid \: i \rangle $ is established
by
\begin{equation}
\ln \mid \langle f\: \mid \: i \rangle \mid\ = 2 \int_0^{\infty}
{d\omega \coth (\omega /2T) \over \omega^2 }\;   {\cal I}m F^R(\omega ,{\bf
R});.
\label{7}
\end{equation}
The effective action (\ref{5},\ref{6}) is correct provided we deal with
heavy particles, $M\gg m_3$.

The formal expression (\ref{6}) is highly nonlinear in $R$ and
can not be solved in general. However in the strong coupling limit,
$g\gg 1$, we can restrict ourselves to a quadratic expansion
\begin{equation}
{\cal F}_n({\bf R})={g \mid \omega_n\mid \over 4\pi}(p_FR)^2 \;,
\label{8}
\end{equation}
which results in a simple quadratic action
\begin{equation}
S^{(2)}={M\beta\over 2}\sum_{n}(\omega^2_n+
\Gamma\mid \omega_n\mid )\mid {\bf R}_n\mid^2\;,
\label{9}
\end{equation}
where ${\bf R}_\tau=\sum_{n}{\bf R}_ne^{i\omega_n\tau}$. Moreover, if we
calculate the mean square value of the particle displacement using Eq.(\ref{9})
\[
p_F^2 \langle\:({\bf R}_{\tau}-{\bf R}_{{\tau}^\prime})^2\:\rangle\approx
{3\over \pi g}\: \ln\:{\Gamma\over 2\pi T}\;;\;\;\;(T\ll \Gamma )\;,
\]
we find \cite{Ya} that the expansion (\ref{8}) is justified in the normal state
down to $T_{coh}$; the higher order
terms in the expansion $
(p_FR)^2-C_4(p_FR)^4+\ldots $ give rise to small corrections
proportional to $(20/ \pi g )C_4\: \ln {\Gamma \over T} $  which can
be neglected
at $T\gg T_{coh}$. The case of negative ions is of most importance here
because for the hard sphere potential with $p_FR_- \gg  1$
($R_-$  is the bubble radius) the coefficient $C_4$
turns out to be very small, $C_4 \sim 10^{-2}\: -\: 10^{-3}$.

Now in the \underline{normal state}, the mobility $\mu $ is given in linear
response, and
in the $R^2$-approximation, by
\begin{equation}
\mu /{\bf e}\:=\: {i\omega \over M\omega^2 - f(\omega )} \equiv {1
\over M}{1 \over  -i\omega +\Gamma } \;,
\label{10}
\end{equation}
where $f(\omega )=\: 4\pi F^R(\omega )/R^2 $, and  ${\bf e}$ is the particle
charge. While this describes diffusive motion for $\omega <\Gamma$, one can
also think of Eqs.(\ref{9},\ref{10}) as describing a frequency-dependent
mass renormalization $M^{eff}/M=1+\Gamma /
\mid \omega_n \mid $. In the diffusive regime one assumes that $M^{eff}<v^2>
\propto T$; then Einstein's relation $\mu \propto <v^2>\tau /T$, with $\tau$
the scattering time, plus the experimentally observed $\mu (T)=const$,
leads to the conclusion that $M^{eff}\propto 1/T $, since certainly
$\tau \ge 1/T $. This however is only very indirect evidence for a
temperature-dependent effective mass.

We therefore consider the ion motion in the \underline{superfluid}
\underline{state}.
To do this we must consider the overlap integral (\ref{7}).
We start from the
expression  derived  by Yamada and Yosida \cite{Yamada82} at $T=0$ for
the normal state which can be readily generalized for the case of
finite temperatures
\begin{equation}
\ln \mid \langle f \! \mid  i \rangle \mid\ = -Tr \int_0^1 \! d\lambda \!
\int {d\omega \over \pi i}(1\!-\!n_\omega ) B^R_{on}(\omega )
A(\omega );
\label{12}
\end{equation}
\begin{eqnarray}
A(\omega )& = & \sum_{n=1}^{\infty}\;
\prod_{i=1}^{n} \; \int {dx_i \over 2\pi i}
 {D(x_1)\ldots D(x_n) \over
(x_1\!-\!\omega \!+\!io)(x_2\!-\!x_1\!+\!io)
\!\ldots\! (x_n\!-\!\omega\! +\!io) }; \nonumber \\
B(\omega )&=&(1-\lambda G(\omega )\Delta \! V )^{-1}\: G(\omega )\Delta \! V
\;; \nonumber \\
D(\omega )&=&-2n_\omega \lambda (1-\lambda G^R(\omega )\Delta \! V )^{-1}\:
 G^R_{on}(\omega )\Delta \! V  \;. \nonumber
\end{eqnarray}
Here $n_\omega $ is the Fermi distribution function, $G(\omega )$ is the Green
function in the initial state, and $\Delta \! V $ is the scattering potential
which distinguishes between the initial and final Hamiltonians, $\Delta \! V
=V_f-V_i$.
We employ the standard definition of the retarded and advanced Green functions
and their on-shell and off-shell parts: $G^{R,A}=G^R_{off}\pm G^R_{on}$;
$G=G^R_{off}+\tanh (\omega /2T) G^R_{on}$. Note, that in the general case
Green functions are matrices not only in the momentum space, but in the
spin and electron-hole channels as well.

Now we make use of the fact that the particle displacement is small
compared to $1/p_F$. This  means  we may treat $\Delta \! V $ as a weak
perturbation
for arbitrary $V$:
\begin{equation}
\Delta \! V _{\bf pp^\prime}=i({\bf p-p^\prime}){\bf R}\:
 (V)_{\bf pp^\prime}\; .
\label{13}
\end{equation}
The leading $R^2$-term in Eq.(\ref{12}) is given by the formula
\[
{-1 \over 2\pi^2}\int\!\!\!\int {dEdE^\prime (1\!-\!n_{E^\prime})n_E \over
(E\!-\!E^\prime \!+\!io)^2}\:Tr \{ G^R_{on}(E)\Delta \! V G^R_{on}(E^\prime
)\Delta \! V  \},
\]
 which can be easily rewritten in the form corresponding to Eq.(\ref{7}). Thus
we can
write the expression for the function $f(\omega )$ (see Eq.(\ref{10})) as
\begin{eqnarray}
&{\cal I}&m f_{\omega }\!= \!np_F \!\int \!\!\!\! \int \! dEdE^\prime
\sigma (E,E^\prime )
(n_{E^\prime}\!-\!n_E)\:\delta (E\!-\!E^\prime \!\!-\!\omega ); \nonumber \\
&\sigma &(E,\! E^\prime )={-1 \over np_F\pi R^2 }\:Tr \{ G^R_{on}(E)\Delta \! V
G^R_{on}(E^\prime )\Delta \! V  \},
\label{15}
\end{eqnarray}
where $n=p_F^3/3\pi^2 $ is the particle density of $^3$He. The static limit of
Eq.(\ref{15})
gives the mobility in the elastic model
 ${\bf e}/\mu =np_F\: \int dE\: \sigma(E,E)\:(-dn_E/dE)$ (see Ref.\cite{Baym}).

Untill now we have not specified the superfluid phase of $^3$He, and
Eqs.(\ref{15}) are valid both for  $^3$He-A and $^3$He-B. To
observe the ballistic motion of ions experimentally we need the scattering
time to be very long ($\tau \ge 1\!-\!10\: \mu s $). This condition under any
reasonable experimental arrangements may be satisfied only in $^3$He-B at
$T\ll T_c$ (in fact $T\le (0.3-0.2) T_c$). So, in the rest of this letter we
concentrate on the effective mass calculation  in $^3$He-B at $T=0$.

It follows from the form of the effective action (\ref{5}) that the mass
renormalization is defined by the $\omega_n^2$-term in the small frequency
expansion of the functional integral
\begin{equation}
\delta \! M=-{2 \over \pi } \int_0^{\infty} {d\omega {\cal I}mf(\omega )
\over \omega^3} \;.
\label{16}
\end{equation}
(in $^3$He-B the effective mass is isotropic). For $T\ll T_c$ we can neglect
the contribution due to normal excitations and further simplify
Eqs.(\ref{15},\ref{16}) to
\begin{equation}
\delta \! M={2np_F R_-^2 \over \Delta }\int \!\!\! \int_{0}^{\infty}
 dxdx^\prime \:
{ {\bar \sigma }(x,-x^\prime ) \over (x+x^\prime )^3}\;.
\label{17}
\end{equation}
where ${\bar\sigma } (x,x^\prime )=\sigma (x,x^\prime )/\pi R_-^2$ depends only
on $(p_FR_-)$, and the dimensionless
frequencies $x=E/\Delta $. In the simplest case of weak scattering potential
(the strong coupling limit $g\gg 1$ still may be realized through a large
number of weak scattering channels contributing to $\sigma $) one can
substitute the Green function in Eq.(\ref{15}) by its unperturbed value, which
is equivalent to performing a $u-v$ Bogoliubov transformation on the
normal state amplitude. After  straightforward algebra we find
\begin{equation}
\delta \! M_{u-v}={\pi \over 16 }\:(1+\sigma_2/2\sigma_{tr})\:{np_F \sigma_{tr}
\over \Delta }\;,
\label{u-v}
\end{equation}
where $\sigma_2=\int d{\bf o}\:(1-\cos ^2 \theta)\: d\sigma/d{\bf o}$.
This mass
renormalization could be as large as  $4\cdot 10^4\: m_3$ at zero pressure.
However, we demonstrate below that the exact calculation for the hard sphere
potential gives a value of $\delta \!M $ substantially different from
Eq.(\ref{u-v}).

 As pointed out
in Ref.\cite{Baym} the scattering matrix has a resonant behavior at
energies near the gap edge which has to be treated exactly. First, we
express conventionally the Green function in terms of the scattering
T-matrix as $G(\omega )=V^{-1}T(\omega )V^{-1}-V^{-1}$ and present the trace
 in Eq.(\ref{15}) in the form
\begin{equation}
Tr \{ T^R_{on}(x)\Delta \! V^{-1} T^R_{on}(x^\prime )\Delta \! V^{-1}  \}\;,
\label{18}
\end{equation}
where $(\Delta \! V^{-1})_{\bf pp^\prime} \equiv i({\bf p-p^\prime}){\bf R}\:
(V^{-1})_{\bf pp^\prime}$. The analytic solution for the T-matrix was found in
Ref.\cite{Baym}. Since the energy spectrum of $^3$He-B is spherically symmetric
and does not depend on spin, and ${\vec \sigma }{\vec p}$ (where ${\vec \sigma
}$
is the fermion spin operator) is invariant under simultaneous rotations of
the spin and momentum, it is clear that T-matrix is diagonal in the total
angular momentum $j$ and its projection $m$. Introducing the angular momentum
eigenstates  $\mid j,m,l \! = \!j\pm 1/2 \rangle \equiv \mid jm\pm \rangle $
one can represent the T-matrix as \cite{Baym,Thun}
\begin{eqnarray}
& & \pi N(0) T={\cal T}\:=\:  \left(
\begin{array}{cc}
 t_1(K)\; & -\sigma_2 t_3(-K)\sigma_2 \\
 t_3(K) \;& \; \sigma_2 t_1(-K)\sigma_2
\end{array} \; \right)\;; \label{19}
 \\
& &t_1=\sum_{jm}\sum_{s=\pm } \mid s> <s\mid \;t_1^{js} \;; \nonumber \\
& &t_3=\sigma_2\: \sum_{jm}\sum_{s=\pm } \mid -s> <s\mid \;t_3^{js} \;,
\nonumber
\end{eqnarray}
where $N(0)$ is the density of states in the normal phase, and
$t_1^{js}$ and $t_3^{js}$ are known functions of frequency and
phase shifts at the Fermi surface,
$K_{j\pm }=\tan \delta_{l=j\pm 1/2} $
\begin{eqnarray}
t_1^{js}&=&K_{js}(1-i\rho K_{j-s})/d_{js} \;; \nonumber \\
t_3^{js}&=&( \rho / x)\: K_{js} K_{j-s} /d_{js} \;; \label{20} \\
d_{js}&=&(1+i\rho K_{js})(1-i\rho K_{j-s})-(\rho /x)^2 K_{js} K_{j-s} \;.
\nonumber
\end{eqnarray}
Here $\rho (x) =x/(x^2-1)^{1/2}$. Obviously, we have the same matrix structure
for  the on-shell retarded T-matrix as that in Eq.(\ref{19}) with the
scattering amplitudes being replaced by the on-shell ones $t \to \tau $
\begin{eqnarray}
\tau_1^{js}&=&  -i\theta (\mid x \mid - 1) {\cal I}m (t_1^{js} )\;;\nonumber \\
\tau_3^{js}&=&  -\theta (\mid x \mid - 1) {\cal R}e (t_3^{js})  \;;
\label{21}
\end{eqnarray}
It is easy to show that the nonzero contribution to the trace comes
from terms having the scattering potential  only in combinations
$1/V_l-1/V_{l^\prime}=\pi N(0) (1/K_l -1/K_{l^\prime})$. So, we may
conveniently replace the scattering potential in Eq.(\ref{18}) with the
K-matrix.
The final expression for $\sigma (x,x^\prime)$ reads
\begin{equation}
\sigma (x,x^\prime )={-1 \over np_F\pi R^2 }Tr \{ {\cal T}^R_{on}(x)\Delta \!
K^{-1} {\cal T}^R_{on}(x^\prime )\Delta \! K^{-1}  \},
\label{22}
\end{equation}

There is one extra contribution to the effective mass due to
bound states in the gap at
$E_{js}=\pm \Delta \: \cos (\delta_{j+} -\delta_{j-})$ \cite{Baym}.
It is proportional to the number
of occupied bound states multiplied by the $^3$He quasiparticle mass
\begin{equation}
\delta \! M_b=m_3^* \sum_{j}^{j_{max}} (2j+1)\;.
\label{23}
\end{equation}
The energies of the bound states approach the gap edge as
 $\Delta -E_j \sim \Delta (d\delta_j/dj)^2/2$. It is physically clear that
states very close to the gap edge will disappear with the ion recoil
in the scattering processes being taken into account. Thus the maximum
orbital number contributing to Eq.(\ref{23}) is defined by $\Delta -E_j \sim
\Delta /g $ (the recoil energy was calculated with the renormalized mass; see
below). For the hard sphere potential with $p_FR_- \gg 1$ the phase shifts
drop abruptly for $j>p_FR_-$, which allows us to estimate the contribution
of the bound states as
\begin{equation}
\delta \! M_b= \approx  (p_FR_-)^2\:m_3^*\approx 3\pi g m_3^*\;.
\label{24}
\end{equation}
This mass is likely to be larger than the bare ion  mass $M$, but still is
much smaller than the renormalization defined by virtual transition in
Eq.(\ref{17}).

The procedure of evaluating $M^{eff} $ is straightforward now, because the
trace determining the function $\sigma (E,E^\prime )$ can be
expressed entirely in terms of phase shifts at the Fermi surface
which for the hard sphere potential are defined as
$ \; \tan \delta_l = j_l(p_FR_-)/n_l(p_FR_-)$, where $j_l$ and $ n_l$ are
the spherical Bessel and Neumann functions of order $l$. For any given pressure
the set of parameters $p_F(P)$, $\Delta (P)$, and $R_-(P)$ allows to
get the effective mass
$M^{eff}=M+\delta \! M_b+\delta \! M$
by numerical evaluation of Eqs.(\ref{17},\ref{22}). In our calculations we used
the normal state parameters taken from Wheatley's tabulation \cite{Wheatley}.
The ion radius was taken from Ref.\cite{Baym}. Unfortunately, we found no
tabulation for $\Delta(T \to 0 ,P)$ in $^3$He-B, and had to rely on a weak
coupling
relation $\Delta(P)=\alpha \cdot  \:1.76\cdot  T_c(P)$ with the
pressure-independent coefficient $\alpha =1.12$ \cite{Nummila}. We think that
$\Delta (P)$ is the
most uncertain parameter in the present calculation. The bare ion mass is
also unknown, but it is unlikely to contribute more than $10 \% $ to the
effective mass, and we simply neglected this contribution.

 Figure 1 shows the effective mass of the negative ion as a function of
pressure. It was found to be as large as $\sim 2\cdot 10^4 m_3$ at zero
pressure
dropping down to $3-4\cdot 10^3 m_3$ for $P>10 \: bar$. The perturbative
result,
Eq.(\ref{u-v}), is shown by the dashed line demonstrating the difference
between
the exact scattering amplitudes in $^3$He-B and those obtained by applying the
$u-v$
transformation on the normal state amplitudes.

The prediction of a very large effective mass should clearly be tested
experimentally. There is already circumstancial evidence for a large mass
in the mobility experiments of Nummila {\it et al.} \cite{Nummila}, who found
that elastic scattering theory seemed to explain their data down to the lowest
temperature obtained in $^3$He-B; this would be hard to understand using a bare
mass assumption, since it would give a recoil energy $\gg T$
already at $T=0.4\:T_c$. However the recoil energy calculated using the
renormalized mass is much smaller, and an elastic scattering model is
valid down to a temperature $T_{el} \approx  p_F^2 / (M \Gamma/\Delta )
 =\Delta / g$.

However what we really require is a \underline{direct} experimental test.
One could search for resonant transitions between the ion energy levels near
the liquid-vapor interface \cite{Shikin}.
The distance between the ion and the surface is large as compared to
the coherence length in $^3$He-B up to the electric field strength $E\sim 100
V/cm$,
with a typical range of resonance frequencies around
$\omega_0 \sim 10-40\: MHz$. From the mobility experiment \cite{Nummila} we
estimate that $\omega_0 \tau \gg 1$ below $0.3\: T_c$, and the resonance
is sharp enough to be observed.
In the time-of-flight experiment the ion mass
in the bulk can be measured for arbitrary pressure. In this technique the
electric field is reversed during a time interval $\Delta t \sim 1-10 \mu s$,
and the ion is supposed not to accelerate above the Landau critical velocity
${\bf e}E \Delta t /M^{eff} \le \Delta /p_F$. Fortunately, the scattering
time is as long as $1\mu s$ already at $0.3\:T_c$, and the enormous ion mass
effectively compensates the smallness of the critical velocity in $^3$He-B.

\acknowledgments

I thank P Stamp and  M Dobroliubov for numerous helpful discussions. I am
also indebted to P Stamp for his constant interest in the work and
valuable comments from which I benefited a lot. This work was supported by
NSERC.

\figure{Effective mass of the negative ion in $^3$He-B. The solid line is the
exact calculation for the hard-sphere potential which is compared to
the  perturbative result (dashed line) with the same transport cross
section in the normal state.}

\end{document}